%% file: ISMIR2020template.tex
\documentclass{article}
\usepackage[T1]{fontenc} 
\usepackage[utf8]{inputenc} 
\usepackage{ismir,amsmath,cite,url}
\usepackage{graphicx}
\graphicspath{./Images/}
\usepackage{color}

\usepackage{import}
\usepackage{xifthen}
\usepackage{pdfpages}
\usepackage{transparent}

\usepackage{tikz}
\usepackage{bbm}
\usetikzlibrary{shapes,arrows}
\usetikzlibrary{positioning}
\tikzstyle{block} = [draw, fill=blue!20, rectangle, 
    minimum height=3em, minimum width=6em]
\tikzstyle{sum} = [draw, fill=blue!20, circle, node distance=1cm]
\tikzstyle{input} = [coordinate]
\tikzstyle{output} = [coordinate]
\tikzstyle{pinstyle} = [pin edge={to-,thin,black}]
\tikzstyle{box} = [draw, thin, fill=blue!20, text width=6em, align = flush center]
\usetikzlibrary{bayesnet}
\usetikzlibrary{arrows}
\usetikzlibrary{calc}
\usepackage{amsfonts}

\usepackage{pgfplots}
\usepackage{pgfplotstable}
\usepackage{color}
\pgfplotsset{compat=1.14}
\pgfmathtruncatemacro{\N}{150}

\usepackage{caption}
\usepackage{subcaption}
\RequirePackage{scrlfile}
\PreventPackageFromLoading{subfig}

\usepackage[bookmarks=false,hidelinks]{hyperref}

\usepackage{lineno}

\title{HpRNet : Incorporating Residual Noise Modeling for Violin in a Variational Parametric Synthesizer}

\oneauthor
{Krishna Subramani, Preeti Rao} {Indian Institute of Technology Bombay \\ {\tt \small subramani.krishna97@gmail.com}}

\begin{document}

\maketitle
\begin{abstract}
Generative Models for Audio Synthesis have been gaining
momentum in the last few years.  More recently, parametric representations of the audio signal have been incorporated to facilitate better musical control of the synthesized output. In this work, we investigate a parametric model for violin tones, in particular the generative modeling of the residual bow noise to make for more natural tone quality. To aid in our analysis, we introduce a dataset of
Carnatic Violin Recordings where bow noise is an integral part of the playing style of higher pitched notes in specific gestural contexts. We obtain insights about each of the harmonic and residual components of the signal, as well as their interdependence,  via observations on the latent space derived in the course of variational encoding of the spectral envelopes of the sustained sounds.
\end{abstract}

\section{Introduction}\label{sec:introduction}
Physical and Spectral Modeling Synthesis are model driven audio modeling procedures. ``Neural Audio Synthesis'' changes the game to that of using data-driven based learning approaches to audio synthesis. Saroff et al. \cite{sarroff2014musical}, Roche et al. \cite{roche2018autoencoders} and Esling et al. \cite{esling2018generative} approached generative synthesis through frame-wise spectral autoencoding, along with additions (like architectural variations, regularization) for more controllable synthesis. Instead of directly modeling the spectrum, Engel et al. \cite{engel2017neural}, Wyse et al. \cite{wyse2018real} and D\'efossez et al. \cite{defossez2018sing} synthesize audio in the time domain, either autoregressively or with RNNs/LSTMS. With the release of the NSynth dataset \cite{engel2017neural} researchers were able to approach synthesis with deep(er) generative models, with the desire to obtain flexible control over the musical attributes like timbre, pitch and loudness.

\begin{figure}[t]
    \centering
    \def\svgwidth{\columnwidth}
    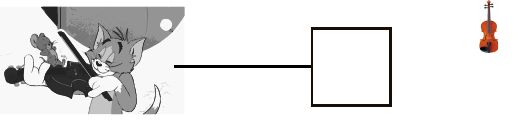

    \vspace{-0.3in}
    \caption{Harmonic plus Residual model}
    \label{fig:hpr_violin}
\end{figure}

Audio can be modelled parametrically in a manner that perceptually relevant parameters become available for musical control over the synthesized sound. A good demonstration of this is the Harmonic plus Residual (HpR) modeling by Serra et al. \cite{serra1989system,serra1997musical} depicted in \autoref{fig:hpr_violin}. The idea is to decompose a signal into a sum of sinusoids whose frequencies are integer multiples of a fundamental frequency, and a residual. Consider the audio signal as $s(t)$,
    \begin{align*} \label{eq::HpR}
        s(t) = \sum_{r=1}^{R} A_{r}(t) \cos(\theta_{r}(t)) + r(t) = h(t) + r(t),
    \end{align*}
where the first term $h(t)$ is the harmonic component, and the second term $r(t)$ is the residual. The residual is in essence that part of the audio signal that cannot be represented by a sum of harmonic partials with $R$ = number of partials used. Examples in musical instruments involve the breathy sound when playing the flute and the the scratchy sound the bow makes when it moves against the violin string during note sustain regions.

The advantage of these parametric models are that they do not require us to model the audio waveform or spectrum directly, rather we can work in the reduced parametric space. Combine this with the generative modeling capabilities of a neural network, and you can obtain a powerful audio synthesizer, one that can rely on small, simple network architectures, can be trained with lesser data, and that can potentially generate high quality audio with musically relevant control over it. Engel et al. \cite{engel2020ddsp} realized this with their Differential Digital Signal Processing pipeline, which used an autoencoder coupled with the HpR model. Subramani et al. \cite{9054181} also combine the same parametric representation with a variational model for controlled synthesis of violin sounds. Neither of the above explicitly considers the modeling of the residual signal.

 \begin{figure*}[t]
    \centering
    \def\svgwidth{\textwidth}
    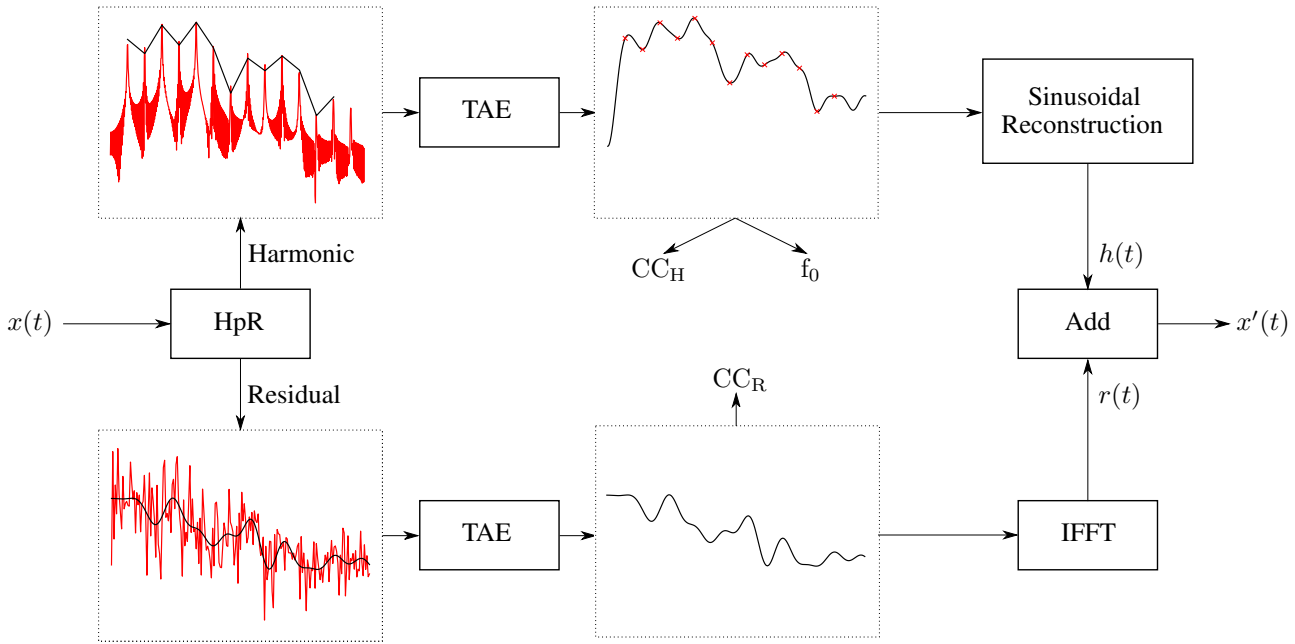
    \caption{Parametric Model for a single frame, Overlap-Add to obtain waveform}
    \label{fig::parammodel}
\end{figure*}

The violin is a popular instrument, both in Western and Indian music. What makes it a popular choice in Carnatic music (classical music from Southern India) is its ability to produce a continuous pitch variation. This is an important component of the melodic motifs of raga music, that involve changing pitch and dynamics throughout the playing gesture. Consider the task of synthesizing a violin solo for a Carnatic music concert. Let us assume we have with us a dataset with a number of notes at different pitches, volumes corresponding to different Carnatic Ragas. Given this, can we train a system for the synthesis of ``natural sounding music'' in the same artist’s style given any ‘musical score’ containing the typical continuous gesture motifs?

Where (or rather how) does parametric modeling come in then? Beauchamp \cite{beauchamp2017comparison} discusses the applicability of the Source-Filter (SF) model to violin audio. Unlike speech, the SF model has not been used widely to model musical instruments because of the possible coupling between the source and filter in instruments \cite{slawson1981color}.  For the violin however, string vibrations are (largely) independent of the body resonances, thus the independence assumption in the SF model is  considered to hold \cite{beauchamp2017comparison,mathews1973electronic}. However, what makes the filter challenging to model is the observation that violin resonances are found  to be much sharper (narrower) than those of voice \cite{beauchamp2017comparison}. This might lead to the indirect dependence of the filter on the source fundamental frequency $\rm{f_0}$ because of the $\rm{f_0}$ dependent sampling of the filter spectral envelope. Subramani et al. \cite{9054181} take care of these `inter-dependencies' between the source and filter by using a conditional variational model which learns the filter conditioned on the source $\rm{f_0}$. However, they only work with the harmonic component and neglect the residual component of the violin, which is a limitation needing to be addressed.
%

Fletcher et al. \cite{fletcher1965quality,fletcher1967quality} performed a very interesting series of experiments on the perceptually important aspects of violin synthesis. The first study \cite{fletcher1965quality} discusses the salient aspects that could differentiate a `real' violin tone from a `synthesized' one. One of those that is of interest to us, and which we will explore further is the residual noise inherent in tone production (the noise produced when drawing the bow across the string). For the lower frequency notes, the fundamental and harmonics mask the noise. However, for the higher frequency notes, they are not able to mask the noise, hence the noise becomes audible. While this `noise' helps in discriminating synthetic notes from real ones, the studies claim that it is usually inaudible for notes of lower frequencies, but becomes audible at notes of higher frequencies. An important thing to consider while synthesizing the violin tone is whether this noise is produced independently of the harmonic component, or whether there are some dependencies. Fletcher et al. in their work synthesize this noise by drawing the bow across the bridge without exciting the strings, thus effectively making it independent of the harmonic component. Mathews et al. \cite{mathews1973electronic} in their studies propose a theory of `Resonant Enhancement' of tones which states that the rich timbre of the violin is essentially due to the string vibrations being filtered at the resonant locations of the violin body. This effectively tells us that if the string vibrations are filtered, then the noise produced by the bowing should also be filtered by the same resonances. Thus, both the harmonic and residual components are produced by the same driving force and cannot be assumed to be independent.
Keeping in mind these possible dependencies between the harmonic and residual components for violin audio, we would like to investigate the joint modeling of the harmonic and residual spectral envelopes. We will do this by evaluating the reconstruction of sustained notes of various pitches and volume dynamics using signal reconstruction error. We also present audio examples of the same in the attached supplementary material.

\section{Parametric Model}\label{sec::bkgrnd}
\autoref{fig::parammodel} summarizes the parametric representation of violin audio that we employ. It is a source-filter inspired representation that builds on top of the HpR model \cite{caetano2012source,caetano2013musical}.
All the blocks mentioned are performed on spectral frames extracted from the sustain portions of single note recordings by applying energy thresholds. 
\begin{enumerate}
    \item We run the HpR model \cite{serra1997musical} on each spectral frame. 
    \item We sub-sample the obtained Harmonic and Residual Spectra. For the Harmonic, we only keep the amplitude peaks corresponding to the harmonic locations, and for the residual, we simply downsample the original spectra to a chosen fixed frequency interval. A residual subsampling rate of 100 Hz is mentioned for a sinusoidal representation of speech in \cite{mcaulay1986speech}. We use a higher subsampling rate of 430 Hz, mentioned by Serra et al. in SMS-Tools \cite{serra1997musical,serra1989system}.
    \item With the sub-sampled spectra, we use the True Amplitude Envelope (TAE) Algorithm \cite{imai_tae_1979,roebel:hal-01161334} to obtain a smooth spectral envelope for each of the harmonic and residual components. The spectral envelopes are represented by their cepstral coefficients. For the harmonic cepstral vector ($\rm{CC_H}$), the number of coefficients is chosen similar to the procedure in \cite{9054181}. The harmonic is also additionally characterized by the fundamental frequency of the frame $\rm{f_0}$. For the residual cepstral vector ($\rm{CC_R}$), we work with a fixed number of cepstral coefficients. 
    \item To reconstruct the harmonic portion, the sinusoid amplitudes are sampled from the harmonic locations of the TAE, and a sinusoidal reconstruction is performed. For the residual, we simply perform the inverse FFT of the residual spectrum with random phases. The net reconstruction is the sum of the two.
\end{enumerate}

We use the HpR model as implemented in SMS-Tools \cite{serra1997musical,serra1989system}. For the TAE algorithm, we use the implementation in \cite{9054181}. 

\section{Dataset}\label{sec::dataset}
There does not exist a publicly available dataset suitable for synthesis of Carnatic Music, especially for the violin. NSynth \cite{engel2017neural} is a large musical note recording dataset. Good-sounds \cite{romani2015real} is also a similar dataset consisting of musical notes and scales recorded for different instruments. However, both of these dataset work with the MIDI notes and are not that expressive. Keeping in mind our task of expressive synthesis, we would ideally like a dataset which is recorded keeping in mind the Carnatic playing style. We recorded an experienced Carnatic violinist playing a set of scale notes at various loudness and playing styles as detailed in the following tables,
\begin{table}[h]
    \centering
    \scalebox{0.8}{
    \begin{tabular}{|l|l|l|l|l|l|l|}
    \hline
    Carnatic Note & $\rm{Sa}$           & $\rm{Ri_{1}}$       & $\rm{Ri_{2}}$        & $\rm{Ga_{2}}$        & $\rm{Ga_{3}}$       & $\rm{Ma_{1}}$       \\ \hline
    Notation      & \textbf{Sa}  & \textbf{Ri1} & \textbf{Ri2}  & \textbf{Ga2}  & \textbf{Ga3} & \textbf{Ma1} \\ \hline
    Carnatic Note & $\rm{Ma_{2}}$       & $\rm{Pa}$           & $\rm{Dha_{1}}$       & $\rm{Dha_{2}}$       & $\rm{Ni_{2}}$       & $\rm{Ni_{3}}$      \\ \hline
    Notation      & \textbf{Ma2} & \textbf{Pa}  & \textbf{Dha1} & \textbf{Dha2} & \textbf{Ni2} & \textbf{Ni3} \\ \hline
    \end{tabular}}
    \caption{Carnatic music notation for the 12 semitones of an octave} \label{tab::carnatic_notes}
\end{table}
\begin{table}[h]
    \centering
    \scalebox{0.8}{
    \begin{tabular}{|l|l|l|}
    \hline
             & Description          & Notation         \\ \hline
    Octave   & Lower, Middle, Upper & \textbf{L, M, U} \\ \hline
    Loudness & Soft, Loud           & \textbf{So, Lo}  \\ \hline
    Style    & Smooth, Attack       & \textbf{Sm, At}  \\ \hline
    \end{tabular}}
    \caption{Recording Parameters} \label{tab::recording_styles}
\end{table}

For each note and choice of style, there are 2 instances recorded, each approximately 2-3 seconds long. More details on the dataset is available in the attached supplementary material. 

\section{Generative Models}
Variational Autoencoders (VAE) \cite{kingma2013auto} are our choice of generative models. They can be viewed as an Autoencoder with a prior enforced on the latent space \cite{doersch2016tutorial}. They minimize the Variational Lower Bound given by,
\begin{equation*}\label{eq::vaeloss}
    \mathcal{L} = \mathbb{E}_{z \sim Q}\{\log P(X|z)\} - \beta D_{KL}\{Q(z|X)||P(z)\},
\end{equation*}
where the first term represents the Mean Squared Error (MSE) between the input and output, and the second term enforces the prior distribution on the latent space. $\beta$ controls the trade-off \cite{higgins2017beta} between the two terms.
A VAE can be thought as an encoder-decoder pair where the encoder outputs the means and variances for the latent distribution. Using the re-parametrization trick \cite{kingma2013auto}, we sample from $\mathcal{N}(\bm{0},\bm{I})$ and transform it through the encoder's mean and variance. This `latent' variable is then passed through the decoder to obtain the network's reconstruction of the input. 

Conditional VAEs \cite{sohn2015learning} work the same way as VAEs, however they condition the input on an additional conditioning variable. Successfully employed in \cite{9054181} for the synthesis of the harmonic component of violin, we extend the same for modeling the residual signal.

\begin{figure}[h]
	\centering
	\begin{subfigure}[h]{\columnwidth}
		\centering
		\includegraphics[width = \columnwidth]{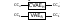}
		\caption{Independent Modeling (INet)}
		\label{fig:exp1_INet}
	\end{subfigure}
	\vspace*{0.2in}
	
	\begin{subfigure}[h]{\columnwidth}
		\centering
		\includegraphics[width = \columnwidth]{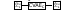}
		\caption{Concatenative Modeling (ConcatNet)}
		\label{fig:exp2_ConcatNet}
	\end{subfigure}

	\vspace{0.2in}
	
	\begin{subfigure}[h]{\columnwidth}
		\centering
		\includegraphics[width = \columnwidth]{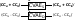}
		\caption{Modeling sum and difference (JNet)}
		\label{fig:exp2_JNet}
	\end{subfigure}
	
	\caption{Network Architectures}
	\label{fig::netarch}
\end{figure}

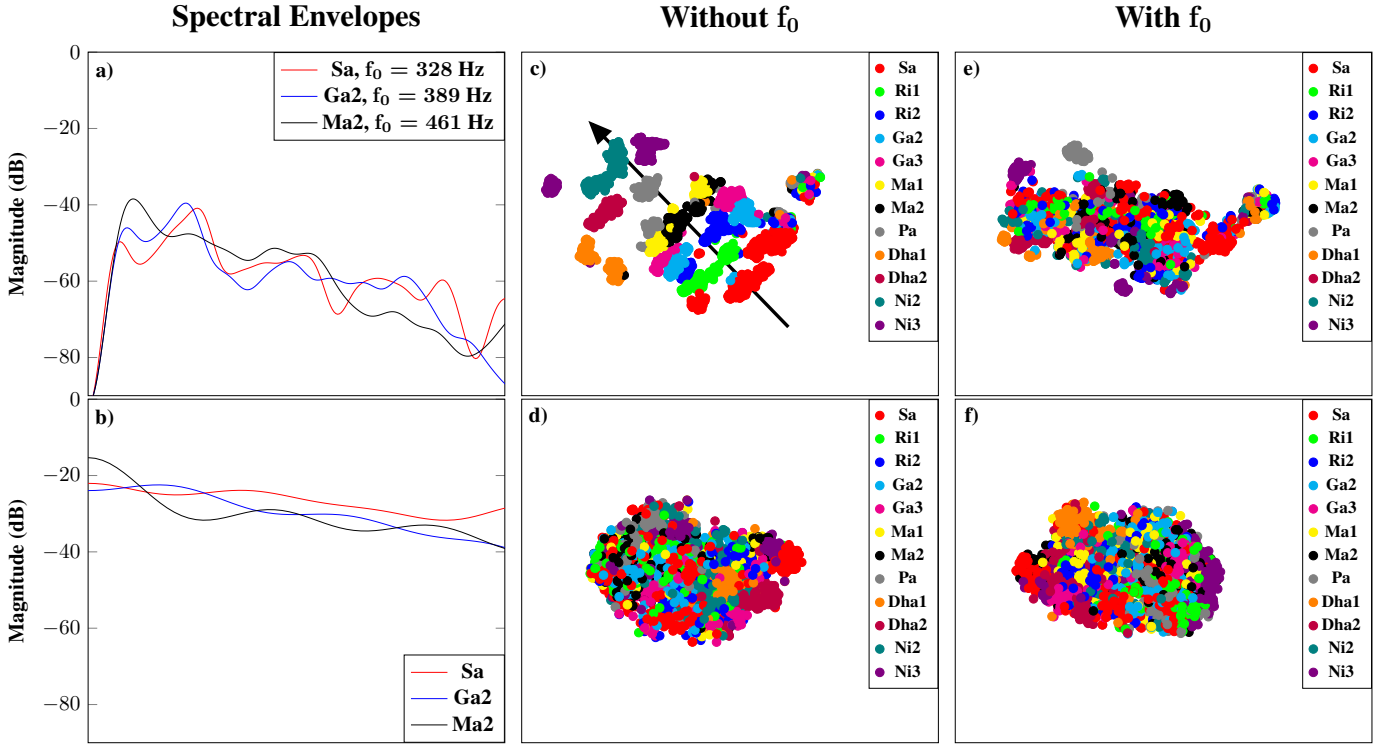
\begin{figure*}[t]
	\centering
	\scalebox{0.8}{
		\mbox{
			\begin{tikzpicture}
			\begin{axis}[title = \Large \textbf{Spectral Envelopes},
			xmajorticks = false,
		 	ylabel={\textbf{Magnitude (dB)}},
			xmin = 0,
			xmax = 5000,
			ymax = -0,
			ymin = -90,
			ylabel style={overlay},
			yticklabel style={overlay},
			xlabel style={overlay},
			xticklabel style={overlay},
			xticklabel = {
				\pgfmathparse{\tick/1000}
				\pgfmathprintnumber[precision = 0]{\pgfmathresult}
			}
			]
			\addplot[color = red] table[col sep=comma,x = f,y = taeH] {./data/02_M_Sa_Sm_So_H.csv};
			\addplot[color = blue] table[col sep=comma,x = f,y = taeH] {./data/08_M_Ga2_Sm_So_H.csv};
			\addplot[color = black] table[col sep=comma,x = f,y = taeH] {./data/17_M_Ma2_Sm_So_H.csv};
			\draw (200,-5) node {\textbf{a)}}; 
			\pgfplotsset{every axis legend/.append style={ at={(1,1)}, anchor=north east}}
			\legend{\textbf{Sa, $\bm{\rm{f_0} = 328}$ Hz},\textbf{Ga2, $\bm{\rm{f_0} = 389}$ Hz},\textbf{Ma2, $\bm{\rm{f_0} = 461}$ Hz}}
			\end{axis}
			\end{tikzpicture}
			
		}   
		\mbox{
			\begin{tikzpicture}
			\begin{axis}[title = \Large \textbf{Without $\bm{\rm{f_0}}$},
			xmajorticks = false,
			ymajorticks = false,
			ymax = 100,
			ymin = -100,
			xmax = 150,
			xmin = -100,
			ylabel style={overlay},
			yticklabel style={overlay},
			xlabel style={overlay},
			xticklabel style={overlay},
			]
			\addplot [scatter,scatter src=explicit symbolic,scatter/classes={0.0={mark=*, color = red}, 1.0={mark=*, color = green}, 2={mark=*, color = blue}, 3={mark=*, color = cyan}, 4={mark=*, color = magenta}, 5={mark=*, color = yellow}, 6={mark=*, color = black}, 7={mark=*, color = gray}, 8={mark=*, color = orange}, 9={mark=*, color = purple}, 10={mark=*, color = teal}, 11={mark=*, color = violet} }, only marks] table[col sep = comma,x = x,y = y, meta = c] {./data/INet_H_644995.csv};
			\pgfplotsset{every axis legend/.append style={ at={(1,1)}, anchor=north east}}
			\draw[color = black, ultra thick, ->] (60,-60) -- (-60,60); 
			\legend{\footnotesize \textbf{Sa},\footnotesize \textbf{Ri1},\footnotesize \textbf{Ri2},\footnotesize \textbf{Ga2},\footnotesize \textbf{Ga3},\footnotesize \textbf{Ma1},\footnotesize \textbf{Ma2},\footnotesize \textbf{Pa},\footnotesize \textbf{Dha1},\footnotesize \textbf{Dha2},\footnotesize \textbf{Ni2},\footnotesize \textbf{Ni3}}
			\draw (-90,89.89) node {\textbf{c)}};
			\end{axis}
			
			\end{tikzpicture}
		}   
		\mbox{
			\begin{tikzpicture}
			\begin{axis}[title = \Large \textbf{With $\bm{\rm{f_0}}$},
			xmajorticks = false,
			ymajorticks = false,
			ymax = 100,
			ymin = -100,
			xmax = 150,
			xmin = -100,
			ylabel style={overlay},
			yticklabel style={overlay},
			xlabel style={overlay},
			xticklabel style={overlay},
			]
			\addplot [scatter,scatter src=explicit symbolic,scatter/classes={0.0={mark=*, color = red}, 1.0={mark=*, color = green}, 2={mark=*, color = blue}, 3={mark=*, color = cyan}, 4={mark=*, color = magenta}, 5={mark=*, color = yellow}, 6={mark=*, color = black}, 7={mark=*, color = gray}, 8={mark=*, color = orange}, 9={mark=*, color = purple}, 10={mark=*, color = teal}, 11={mark=*, color = violet} }, only marks] table[col sep = comma,x = x,y = y, meta = c] {./data/INet_H_342604.csv};
			
			\pgfplotsset{every axis legend/.append style={ at={(1,1)}, anchor=north east}}
			\legend{\footnotesize \textbf{Sa},\footnotesize \textbf{Ri1},\footnotesize \textbf{Ri2},\footnotesize \textbf{Ga2},\footnotesize \textbf{Ga3},\footnotesize \textbf{Ma1},\footnotesize \textbf{Ma2},\footnotesize \textbf{Pa},\footnotesize \textbf{Dha1},\footnotesize \textbf{Dha2},\footnotesize \textbf{Ni2},\footnotesize \textbf{Ni3}}
			\draw (-90,89.89) node {\textbf{e)}};
			\end{axis}
			
			\end{tikzpicture}
		}
	}
	\scalebox{0.8}{
		\mbox{
			\begin{tikzpicture}
			\begin{axis}[
			xmajorticks = false,
			ylabel={\textbf{Magnitude (dB)}},
			xmin = 0,
			xmax = 5000,
			ymax = -0,
			ymin = -90,
			ylabel style={overlay},
			yticklabel style={overlay},
			xlabel style={overlay},
			xticklabel style={overlay},
			xticklabel = {
				\pgfmathparse{\tick/1000}
				\pgfmathprintnumber[precision = 0]{\pgfmathresult}
			}
			]
			\addplot[color = red] table[col sep=comma,x = f,y = taeR] {./data/02_M_Sa_Sm_So_R.csv};
			\addplot[color = blue] table[col sep=comma,x = f,y = taeR] {./data/08_M_Ga2_Sm_So_R.csv};
			\addplot[color = black] table[col sep=comma,x = f,y = taeR] {./data/17_M_Ma2_Sm_So_R.csv};
			\draw (200,-5) node {\textbf{b)}}; 
			\pgfplotsset{every axis legend/.append style={ at={(1,0)}, anchor=south east}}
			\legend{\textbf{Sa},\textbf{Ga2},\textbf{Ma2}}
			\end{axis}
			\end{tikzpicture}
			
		}   
		\mbox{
			\begin{tikzpicture}
			\begin{axis}[
			xmajorticks = false,
			ymajorticks = false,
			ymax = 100,
			ymin = -100,
			xmax = 150,
			xmin = -100,
			ylabel style={overlay},
			yticklabel style={overlay},
			xlabel style={overlay},
			xticklabel style={overlay},
			]
			\addplot [scatter,scatter src=explicit symbolic,scatter/classes={0.0={mark=*, color = red}, 1.0={mark=*, color = green}, 2={mark=*, color = blue}, 3={mark=*, color = cyan}, 4={mark=*, color = magenta}, 5={mark=*, color = yellow}, 6={mark=*, color = black}, 7={mark=*, color = gray}, 8={mark=*, color = orange}, 9={mark=*, color = purple}, 10={mark=*, color = teal}, 11={mark=*, color = violet} }, only marks] table[col sep = comma,x = x,y = y, meta = c] {./data/INet_R_850649.csv};
			\pgfplotsset{every axis legend/.append style={ at={(1,1)}, anchor=north east}}
			\legend{\footnotesize \textbf{Sa},\footnotesize \textbf{Ri1},\footnotesize \textbf{Ri2},\footnotesize \textbf{Ga2},\footnotesize \textbf{Ga3},\footnotesize \textbf{Ma1},\footnotesize \textbf{Ma2},\footnotesize \textbf{Pa},\footnotesize \textbf{Dha1},\footnotesize \textbf{Dha2},\footnotesize \textbf{Ni2},\footnotesize \textbf{Ni3}}
			\draw (-90,89.89) node {\textbf{d)}};
			\end{axis}
			
			\end{tikzpicture}
		}   
		\mbox{
			\begin{tikzpicture}
			\begin{axis}[
			xmajorticks = false,
			ymajorticks = false,
			ymax = 100,
			ymin = -100,
			xmax = 150,
			xmin = -100,
			ylabel style={overlay},
			yticklabel style={overlay},
			xlabel style={overlay},
			xticklabel style={overlay},
			]
			\addplot [scatter,scatter src=explicit symbolic,scatter/classes={0.0={mark=*, color = red}, 1.0={mark=*, color = green}, 2={mark=*, color = blue}, 3={mark=*, color = cyan}, 4={mark=*, color = magenta}, 5={mark=*, color = yellow}, 6={mark=*, color = black}, 7={mark=*, color = gray}, 8={mark=*, color = orange}, 9={mark=*, color = purple}, 10={mark=*, color = teal}, 11={mark=*, color = violet} }, only marks] table[col sep = comma,x = x,y = y, meta = c] {./data/INet_R_565174.csv};
			
			\pgfplotsset{every axis legend/.append style={ at={(1,1)}, anchor=north east}}
			\legend{\footnotesize \textbf{Sa},\footnotesize \textbf{Ri1},\footnotesize \textbf{Ri2},\footnotesize \textbf{Ga2},\footnotesize \textbf{Ga3},\footnotesize \textbf{Ma1},\footnotesize \textbf{Ma2},\footnotesize \textbf{Pa},\footnotesize \textbf{Dha1},\footnotesize \textbf{Dha2},\footnotesize \textbf{Ni2},\footnotesize \textbf{Ni3}}
			\draw (-90,89.89) node {\textbf{f)}};
			\end{axis}
			
			\end{tikzpicture}
		}   
	}
	\caption{Spectral Envelopes, Latent Space visualizations for Harmonic (top), Residual (bottom)}
	\label{fig::exp1}
\end{figure*}

\section{Network Architecture}
The inputs to our CVAEs are the harmonic or residual CCs ($\rm{CC_H}$, $\rm{CC_R}$), along with pitch $\rm{f_0}$ as a conditional input for the harmonic CCs. We follow the experimental procedure presented in \cite{9054181} to obtain the optimal values of the hyperparameters. The values are $\beta = 10^{-3}$ and latent space dimensionality of 32.  
The Encoder is a linear fully connected neural network with leaky ReLU activations (allows for stable training and the output to have negative values). The Decoder is the same architecture as the encoder, but with reversed dimensions. We have implemented all the networks in PyTorch \cite{paszke2017automatic}. We train our model on a mobile Nvidia GeForce GTX 1070 with batch size 512. We use ADAM \cite{kingma2014adam} as the optimizer with an initial learning rate of $10^{-3}$, and run the optimization for 2000 epochs.

\section{Experiments}
We investigate the following two aspects in our work,
\begin{enumerate}
    \item The role of explicit pitch conditioning in the VAE to model the harmonic and residual components.
    \item Possible inter-dependencies between the harmonic and residual components.
\end{enumerate}
\autoref{fig::netarch} summarizes the 3 network architectures we investigate.
For the first task, we will independently model the harmonic and residual components with individual CVAEs as shown in \autoref{fig:exp1_INet}, and use a combination of spectral envelop plots and CVAE latent space visualizations to obtain insights. For the second task, we shall introduce networks that jointly model the harmonic and residual components as shown in \autoref{fig:exp2_ConcatNet}, \autoref{fig:exp2_JNet}.
    
\subsection{Pitch Conditioning}
The traditional SF model from speech processing assumes independence between the source and filter, which is largely true for vocal apparatus. If that were the case for the violin as well, then we should in principle be able to model the violin by only modeling the spectral envelope for a single $\rm{f_0}$. 

\autoref{fig::exp1} a) shows Spectral Envelopes for different $\rm{f_0}$. The shape differs across pitches. However, rather than these variations occurring because of the non-independence of the source and filter, we speculate that these occur because of the narrow resonances in the violin body \cite{beauchamp2017comparison}. Thus, even for a slight change in $\rm{f_0}$, the relative amplitudes can change quite drastically. This has been noted by Beauchamp in \cite{beauchamp2017comparison} and Fletcher in \cite{fletcher1967quality}. The envelopes we plot in \autoref{fig::exp1} a) show exactly this variation across pitches. Thus, by conditioning the envelopes on the pitch, we can expect the network to better reconstruct the spectral envelope.

To further convince ourselves of the need for conditioning, we visualize the latent spaces of our VAE. \autoref{fig:exp1_INet} shows the network we employ. Since our latent space is quite high dimensional (32 in our case), to visualize it, we use the t-SNE algorithm \cite{maaten2008visualizing} that projects high dimensional data onto lower dimensions (2 in our case), and helps in effectively visualizing clusters in the data. \autoref{fig::exp1} c), e) shows the harmonic latent spaces without and with pitch conditioning. If the harmonic spectral envelope was independent of pitch, then we should ideally not be seeing any clustering in the latent space. However, we can see considerible clustering when we do not condition on the pitch. Another interesting thing to observe in the clustering is its structure. For close notes, the clusters are close, and the clusters move away (from right to left) as you progress from the Sa to Ni3. The black arrow overlaid on top shows the progression of note clusters from Sa to Ni3.
In essence, this plot tells us that the latent space still contains information on the pitch, thus providing additional motivation to condition the envelope on the pitch. On doing this, we can see in the latent space that all the notes are clustered around together. Thus, with the pitch as a conditional, the decoder can correctly sample the latent space to obtain the correct harmonic envelope for that pitch.

The Residual envelopes depict a different picture though. \autoref{fig::exp1} b) shows that the Residual Spectral envelope does not significantly change for different pitches, thus hinting that the residual spectral envelope is indeed not dependent on the pitch (as the SF model suggests). This can be explained by the fact that we have sufficiently sub-sampled the actual residual spectrum to capture variations in the envelope. The residual latent space visualizations \autoref{fig::exp1} d), f) also re-affirm our conclusion. You do not observe any kind of clustering, either without or with pitch conditioning, thus suggesting that the residual spectral envelopes are indeed independent of the pitch.


\subsection{Interdependence of Harmonic, Residual}
\begin{figure}[h]
	\centering
	    \begin{minipage}[h]{0.2\textwidth}
	    \centering
	    \scalebox{0.45}{
		    \begin{tikzpicture}
        \begin{axis}[title = \Large \textbf{Harmonic},
        	xlabel={\Large \textbf{Frequency (kHz)}},
        	ylabel={\Large \textbf{Magnitude (dB)}},
        	xmin = 0,
        	xmax = 5000,
        	ymax = 0,
        	ymin = -100,
        	xticklabel style = {font=\Large \boldmath},
        	yticklabel style = {font=\Large \boldmath},
         	xticklabel = {
             \pgfmathparse{\tick/1000}
             \pgfmathprintnumber{\pgfmathresult}
         	},
        ylabel style={overlay},
                    yticklabel style={overlay},
                    xticklabel style={overlay},
            ]
            \addplot[color = red,thick] table[col sep=comma,x = f,y = taeH] {./data/19_M_Ma2_Sm_So_H.csv};
            \addplot[color = black,thick] table[col sep=comma,x = f,y = taeH] {./data/255_M_Ma2_Sm_Lo_H.csv};
            \legend{\Large \textbf{Soft},\Large \textbf{Loud}};
            \addplot[mark = *o, only marks] table[col sep=comma,x = nF,y = nM] {./data/19_M_Ma2_Sm_So_H_harmlocs.csv}
            \foreach \i in {0,...,\N} {
                        coordinate [pos=\i/\N] (a\i)
                    }
            ;
            
            \addplot[mark = *o, only marks] table[col sep=comma,x = nF,y = nM] {./data/255_M_Ma2_Sm_Lo_H_harmlocs.csv}
            \foreach \i in {0,...,\N} {
                        coordinate [pos=\i/\N] (b\i)
                    }
            ;
        
        
            
            \end{axis}
            \foreach \i in {0,...,9} {
                    \draw[color = blue, thick] (a\i) -- (b\i);
                }
            \end{tikzpicture}}
	    \end{minipage}
	    \begin{minipage}[h]{0.2\textwidth}
	    \centering
	    \scalebox{0.45}{
		    \begin{tikzpicture}
            \begin{axis}[title = \Large \textbf{Residual},
            	x filter/.code={\pgfmathparse{#1/1000}\pgfmathresult},
                xlabel={\Large \textbf{Frequency (kHz)}},
                xmin = 0,
                xmax = 20,
                ymax = 0,
                ymin = -100,
                xticklabel style = {font=\Large \boldmath},
                yticklabel style = {font=\Large \boldmath},
                tick label style={
            font= \boldmath},
        	xtick={0,5,10,15,20},
            ylabel style={overlay},
                    yticklabel style={overlay},
                    xticklabel style={overlay},
            ]
            \addplot[color = red,thick] table[col sep=comma,x = f,y = taeR] {./data/19_M_Ma2_Sm_So_R.csv};
            \addplot[color = black,thick] table[col sep=comma,x = f,y = taeR] {./data/255_M_Ma2_Sm_Lo_R.csv};
            \legend{\Large \textbf{Soft},\Large \textbf{Loud}};
        
            
            \end{axis}
            \end{tikzpicture}}
	    \end{minipage}
	    
    \caption{Spectral Envelopes for Soft and Loud notes}
    \label{fig::exp2-1}
\end{figure}
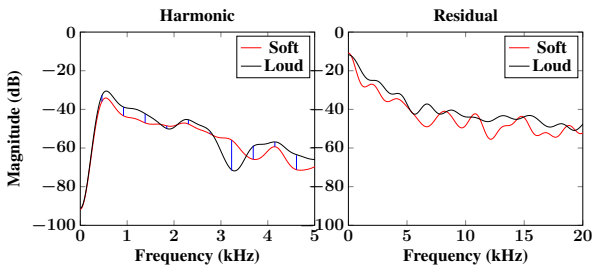

One important question still remains. Are the harmonic and residual portions somehow coupled to each other? If this is the case, simply modeling the individual components with independent networks could be sub-optimal. Where do these dependencies arise from - to answer that, we go back to the `Resonant Enhancement' theory of tones \cite{mathews1973electronic}   which states that the rich timbre of the violin is essentially due to the string vibrations being filtered at the resonant locations of the violin body. When we bow the string harder to produce a louder tone, the residual component will also be loud, and they both will be filtered by the violin body simultaneously, thus indicating that the harmonic and residual fundamentally depend on the playing style of the note. To check our hypothesis, we show the harmonic and residual spectral envelope variations in \autoref{fig::exp2-1} for the same note by varying the loudness from soft to loud.

The blue lines in the harmonic spectral envelope in \autoref{fig::exp2-1} represent the magnitude differences for the harmonics. If loudness variation were a simple amplitude scaling, then both the harmonic and spectral envelopes should be shifted up (log-plots) and the blue lines should all be the same length. However, as we see, a loudness increase is not just a scaling. It causes certain frequencies to be boosted, others to be suppressed, and also changes the tilt in the spectral envelope. This further strengthens our hypothesis that the harmonic and residual envelopes must be dependent as they have a common underlying origin in the played style of the note. \autoref{fig:exp2_ConcatNet}, \autoref{fig:exp2_JNet} shows the 2 additional network architectures we try out, besides the independent modeling used in the first experiment.

There could be many different ways to try joint modeling in a neural network. The simplest procedure however is to simply concatenate the inputs and feed them to a CVAE to model them together, as shown in \autoref{fig:exp2_ConcatNet}. Since the encoder and decoder are given as input both the harmonic and residual CCs, the reconstruction inherently takes into account both the harmonic and residual components.
The second approach of modeling the sum and difference of CCs is more non-trivial. The intuition behind it comes from current methods that generatively model the magnitude spectrum of the sound \cite{sarroff2014musical,roche2018autoencoders}. The magnitude spectra is the sum of the harmonic and residual spectra. Thus, by directly modeling the spectrum, the autoencoder takes care of both of them together. If we could somehow model the difference of the harmonic and residual spectra as well, we could individually obtain the harmonic and residual components. That is exactly what we try to do via our network, as shown \autoref{fig:exp2_JNet}. We have 2 networks, the sum and difference networks. The sum network, in the process of autoencoding the sum  of the harmonics and residual inherently learns their joint dependencies. The difference network is a `trick' to extract the individual harmonic and residual components from the sum network. We can obtain the harmonic and residual vectors by simply adding and subtracting the outputs of the sum and difference networks. One might ask why do we need the individual components? Keeping in mind the end-goal of being able to synthesize audio, it would be good to have the harmonic and residual components if one is additionally interested in `modifying' the audio (time stretching, frequency scaling, morphing etc.)

How to decide which network works better? We plot the reconstruction MSE, which is computed as the average over all test instance frames given as input to the network (test here refers to the fact that the network has not seen these during training).
We work with the sustain portion of the notes in our dataset, and split it to train and test data evenly. To allow the network to learn the potential dependencies of the harmonic and residual components, we train with frames of both loudness's - soft and loud. Also, we choose notes in the higher octave because Fletcher et al. \cite{fletcher1965quality,fletcher1967quality} mentions explicitly that the residual plays a more important role perceptually in the higher octaves. Thus, with this joint modeling, we hope to see the residual being reconstructed at a lower MSE.

\newcommand{\ymin}{0e-2}
\newcommand{\ymax}{5e-2}
\newcommand{\msela}{040501} 
\newcommand{\mselb}{734506} 
\newcommand{\mselc}{133514} 
\begin{figure}[h]
	\centering
	\begin{minipage}[h]{0.2\textwidth}
		\centering
		\scalebox{0.45}{
			\begin{tikzpicture}
			\begin{axis}[title = \Large \textbf{Harmonic MSE},
			xlabel={\normalsize \Large \textbf{Note}},
			ylabel={\normalsize \Large \textbf{MSE}},
			xtick={0,1,2,3,4},
			xticklabels = {Sa,Ri1,Ri2,Ga2,Ga3},
			 xticklabel style = {font=\Large \boldmath},
			 yticklabel style = {font=\Large \boldmath},
			grid=both,
			ymax = \ymax,
			ymin = \ymin,
			xmin = 0,
			xmax = 4,
			ylabel style={overlay},
			yticklabel style={overlay},
			xticklabel style={overlay},
			]
			
			\addplot [color = red] table[col sep = comma,x expr=\coordindex,y = \msela] {./data/dump-csv/INet_H.csv};
			\addplot [color = blue] table[col sep = comma,x expr=\coordindex,y = \mselc] {./data/dump-csv/HpRNet_H.csv};
			\addplot [color = black] table[col sep = comma,x expr=\coordindex,y = \mselb] {./data/dump-csv/ConcatNet_H.csv};
			
			
			\pgfplotsset{every axis legend/.append style={ at={(0,1)}, anchor=north west}};
			\legend{\Large \textbf{INet}, \Large \textbf{JNet}, \Large \textbf{ConcatNet}};
			\end{axis}
			
			\end{tikzpicture}}
	\end{minipage}
	\begin{minipage}[h]{0.2\textwidth}
		\centering
		\scalebox{0.45}{
			\begin{tikzpicture}
			\begin{axis}[title = \Large \textbf{Residual MSE},
				xlabel={\normalsize \Large \textbf{Note}},
				xtick={0,1,2,3,4},
				xticklabels = {Sa,Ri1,Ri2,Ga2,Ga3},
				 xticklabel style = {font=\Large \boldmath},
				 yticklabel style = {font=\Large \boldmath},
				grid=both,
				ymax = \ymax,
				ymin = \ymin,
				xmin = 0,
				xmax = 4,
				ylabel style={overlay},
				yticklabel style={overlay},
				xticklabel style={overlay},
				]
				
				\addplot [color = red] table[col sep = comma,x expr=\coordindex,y = \msela] {./data/dump-csv/INet_R.csv};
				\addplot [color = blue] table[col sep = comma,x expr=\coordindex,y = \mselc] {./data/dump-csv/HpRNet_R.csv};
				\addplot [color = black] table[col sep = comma,x expr=\coordindex,y = \mselb] {./data/dump-csv/ConcatNet_R.csv};
				
				
				\pgfplotsset{every axis legend/.append style={ at={(0,1)}, anchor=north west}};
				\legend{\Large \textbf{INet}, \Large \textbf{JNet}, \Large \textbf{ConcatNet}};
			\end{axis}
			
		\end{tikzpicture}}
	\end{minipage}
	
	\caption{Reconstruction MSE}
	\label{fig::exp2-2}
\end{figure}
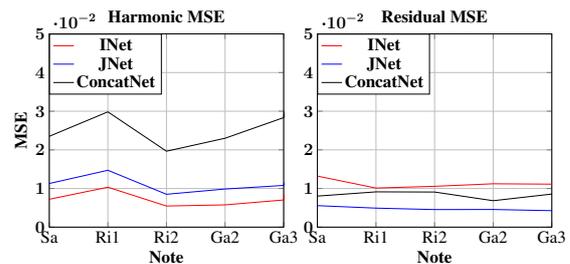
\newpage
\autoref{fig::exp2-2} shows the note-wise reconstruction MSE. For the Harmonic MSE, the Independent Modeling proves to be most superior. Interestingly, for the Residual MSE, the joint modeling methods result in a lower MSE though, thus strengthening our belief in joint modeling of the harmonic and residual components. We have also presented a few audio examples of note reconstructions in the attached supplementary material.

\section{Conclusion}
We introduce HpRNet - a framework combining generative synthesis with parametric modeling of audio. To aid in our analysis, we also introduce a new Carnatic Violin dataset, which we plan on making open to the MIR community. We highlight the necessity of pitch conditioning for the harmonic component. We also provide motivation to jointly model the harmonic and residual components instead of independently modeling them. The reconstruction MSE plots only give us a partial picture - to analyze the perceptual aspects of the reconstructed notes, we plan to conduct listening tests in the future where we present the outputs from our model to experienced Carnatic violinists, and ask them to rate how `good' they think the sound is, which will help us in zeroing onto the perceptually relevant aspects of the audio for synthesis. This work modeled the sustain regions frame wise; the attack needs to be modeled for a complete representation. Eventually, we hope to apply this work to the synthesis of natural sounding raga motifs or ornaments on the violin, characterized as they are by specific pitch and loudness dynamics.
We also hope that our dataset encourages further research in Carnatic music synthesis.

\bibliography{ISMIRtemplate}

\end{document}

%% file: Images/HpR.pdf_tex
\begingroup%
  \makeatletter%
  \providecommand\color[2][]{%
    \errmessage{(Inkscape) Color is used for the text in Inkscape, but the package 'color.sty' is not loaded}%
    \renewcommand\color[2][]{}%
  }%
  \providecommand\transparent[1]{%
    \errmessage{(Inkscape) Transparency is used (non-zero) for the text in Inkscape, but the package 'transparent.sty' is not loaded}%
    \renewcommand\transparent[1]{}%
  }%
  \providecommand\rotatebox[2]{#2}%
  \newcommand*\fsize{\dimexpr\f@size pt\relax}%
  \newcommand*\lineheight[1]{\fontsize{\fsize}{#1\fsize}\selectfont}%
  \ifx\svgwidth\undefined%
    \setlength{\unitlength}{151.05617523bp}%
    \ifx\svgscale\undefined%
      \relax%
    \else%
      \setlength{\unitlength}{\unitlength * \real{\svgscale}}%
    \fi%
  \else%
    \setlength{\unitlength}{\svgwidth}%
  \fi%
  \global\let\svgwidth\undefined%
  \global\let\svgscale\undefined%
  \makeatother%
  \begin{picture}(1,0.25336827)%
    \lineheight{1}%
    \setlength\tabcolsep{0pt}%
    \put(0,0){\includegraphics[width=\unitlength,page=1]{HpR.pdf}}%
    \put(0.3814,0.1350763){\color[rgb]{0,0,0}\makebox(0,0)[lt]{\lineheight{1.25}\smash{\begin{tabular}[t]{l}$h(t) + r(t)$\end{tabular}}}}%
    \put(0.6285002,0.12224792){\color[rgb]{0,0,0}\makebox(0,0)[lt]{\lineheight{1.25}\smash{\begin{tabular}[t]{l}\bf HpR\end{tabular}}}}%
    \put(0,0){\includegraphics[width=\unitlength,page=2]{HpR.pdf}}%
    \put(0.79507524,0.16272508){\color[rgb]{0,0,0}\rotatebox{30.272119}{\makebox(0,0)[lt]{\lineheight{1.25}\smash{\begin{tabular}[t]{l}$h(t)$\end{tabular}}}}}%
    \put(0.78884492,0.06807056){\color[rgb]{0,0,0}\rotatebox{-29.727881}{\makebox(0,0)[lt]{\lineheight{1.25}\smash{\begin{tabular}[t]{l}$r(t)$\end{tabular}}}}}%
  \end{picture}%
\endgroup%

%% file: Images/HpR_flow.pdf_tex
\begingroup%
  \makeatletter%
  \providecommand\color[2][]{%
    \errmessage{(Inkscape) Color is used for the text in Inkscape, but the package 'color.sty' is not loaded}%
    \renewcommand\color[2][]{}%
  }%
  \providecommand\transparent[1]{%
    \errmessage{(Inkscape) Transparency is used (non-zero) for the text in Inkscape, but the package 'transparent.sty' is not loaded}%
    \renewcommand\transparent[1]{}%
  }%
  \providecommand\rotatebox[2]{#2}%
  \newcommand*\fsize{\dimexpr\f@size pt\relax}%
  \newcommand*\lineheight[1]{\fontsize{\fsize}{#1\fsize}\selectfont}%
  \ifx\svgwidth\undefined%
    \setlength{\unitlength}{208.96588898bp}%
    \ifx\svgscale\undefined%
      \relax%
    \else%
      \setlength{\unitlength}{\unitlength * \real{\svgscale}}%
    \fi%
  \else%
    \setlength{\unitlength}{\svgwidth}%
  \fi%
  \global\let\svgwidth\undefined%
  \global\let\svgscale\undefined%
  \makeatother%
  \begin{picture}(1,0.48959387)%
    \lineheight{1}%
    \setlength\tabcolsep{0pt}%
    \put(0,0){\includegraphics[width=\unitlength,page=1]{HpR_flow.pdf}}%
    \put(0.15634747,0.24011205){\color[rgb]{0,0,0}\makebox(0,0)[lt]{\lineheight{1.25}\smash{\begin{tabular}[t]{l}HpR\end{tabular}}}}%
    \put(0,0){\includegraphics[width=\unitlength,page=2]{HpR_flow.pdf}}%
    \put(0.34620894,0.40341536){\color[rgb]{0,0,0}\makebox(0,0)[lt]{\lineheight{1.25}\smash{\begin{tabular}[t]{l}TAE\end{tabular}}}}%
    \put(0,0){\includegraphics[width=\unitlength,page=3]{HpR_flow.pdf}}%
    \put(0.34620894,0.07752335){\color[rgb]{0,0,0}\makebox(0,0)[lt]{\lineheight{1.25}\smash{\begin{tabular}[t]{l}TAE\end{tabular}}}}%
    \put(-0.00157304,0.24011205){\color[rgb]{0,0,0}\makebox(0,0)[lt]{\lineheight{1.25}\smash{\begin{tabular}[t]{l}$x(t)$\end{tabular}}}}%
    \put(0,0){\includegraphics[width=\unitlength,page=4]{HpR_flow.pdf}}%
    \put(0.47553448,0.28133881){\color[rgb]{0,0,0}\makebox(0,0)[lt]{\lineheight{1.25}\smash{\begin{tabular}[t]{l}$\rm{CC_H}$\end{tabular}}}}%
    \put(0.60534514,0.28111192){\color[rgb]{0,0,0}\makebox(0,0)[lt]{\lineheight{1.25}\smash{\begin{tabular}[t]{l}$\rm{f_0}$\end{tabular}}}}%
    \put(0.53783046,0.19646587){\color[rgb]{0,0,0}\makebox(0,0)[lt]{\lineheight{1.25}\smash{\begin{tabular}[t]{l}$\rm{CC_R}$\end{tabular}}}}%
    \put(0,0){\includegraphics[width=\unitlength,page=5]{HpR_flow.pdf}}%
    \put(0.7792275,0.41375235){\color[rgb]{0,0,0}\makebox(0,0)[lt]{\lineheight{1.25}\smash{\begin{tabular}[t]{l}Sinusoidal\end{tabular}}}}%
    \put(0.75895736,0.39200256){\color[rgb]{0,0,0}\makebox(0,0)[lt]{\lineheight{1.25}\smash{\begin{tabular}[t]{l}Reconstruction\end{tabular}}}}%
    \put(0,0){\includegraphics[width=\unitlength,page=6]{HpR_flow.pdf}}%
    \put(0.80457291,0.07752335){\color[rgb]{0,0,0}\makebox(0,0)[lt]{\lineheight{1.25}\smash{\begin{tabular}[t]{l}IFFT\end{tabular}}}}%
    \put(0,0){\includegraphics[width=\unitlength,page=7]{HpR_flow.pdf}}%
    \put(0.80731183,0.24011205){\color[rgb]{0,0,0}\makebox(0,0)[lt]{\lineheight{1.25}\smash{\begin{tabular}[t]{l}Add\end{tabular}}}}%
    \put(0.93877185,0.24011205){\color[rgb]{0,0,0}\makebox(0,0)[lt]{\lineheight{1.25}\smash{\begin{tabular}[t]{l}$x'(t)$\end{tabular}}}}%
    \put(0,0){\includegraphics[width=\unitlength,page=8]{HpR_flow.pdf}}%
    \put(0.18254681,0.29251034){\color[rgb]{0,0,0}\makebox(0,0)[lt]{\lineheight{1.25}\smash{\begin{tabular}[t]{l}Harmonic\end{tabular}}}}%
    \put(0.18218774,0.18375998){\color[rgb]{0,0,0}\makebox(0,0)[lt]{\lineheight{1.25}\smash{\begin{tabular}[t]{l}Residual\end{tabular}}}}%
    \put(0.83329378,0.29251034){\color[rgb]{0,0,0}\makebox(0,0)[lt]{\lineheight{1.25}\smash{\begin{tabular}[t]{l}$h(t)$\end{tabular}}}}%
    \put(0.8331194,0.18375998){\color[rgb]{0,0,0}\makebox(0,0)[lt]{\lineheight{1.25}\smash{\begin{tabular}[t]{l}$r(t)$\end{tabular}}}}%
    \put(0,0){\includegraphics[width=\unitlength,page=9]{HpR_flow.pdf}}%
  \end{picture}%
\endgroup%